\documentclass[prl,aps,preprint,showpacs,showkeys]{revtex4}
\begin{document}
\date{\today}
\title{Families and clustering in a natural numbers network}

\author{Gilberto Corso }

\affiliation{ International Center for Complex Systems and
 Departamento de Biof{\'\i}sica, 
    Centro de Bioci\^encias, 
Universidade Federal do Rio Grande do Norte, Campus Universit\'ario
 59072 970, Natal, RN, Brazil.}

\begin{abstract}

We develop a network in which the natural numbers are the vertices. The 
decomposition of natural numbers by prime numbers is used to establish the 
connections.  We perform data collapse and show that the degree 
distribution of these networks  scales  
linearly with the number of vertices. We explore th families of vertices 
in connection with prime numbers decomposition.  We compare the average 
distance of the network and the clustering coefficient with the distance 
and clustering coefficient of the corresponding random graph. In case 
we set connections among vertices each time the numbers share a common 
prime number the network has properties similar to a random graph. 
If the criterium for establishing 
links becomes more selective, only prime numbers greater 
than $p_l$ are used to establish links, the network   
 has high clustering coefficient.

\end{abstract}


\maketitle

\section{I. Introduction}
The new century have started with a strong development in 
network theory and specially in small-world networks. 
Three ingredients are necessary to define  
a small-world network: a sparse network, small distance and  
high clustering coefficient \cite{WTmodel,barabasi}. Many examples of these networks 
have been analyzed in diverse fields as  
computation \cite{computation1,computation2},
linguistic \cite{linguistic1,linguistic2}, biology \cite{biology1,biology2}, 
economics \cite{economics,eco2} and social phenomena \cite{new1,sex}. 
In addition to the interest in the description of these particular 
phenomena, small-world networks are promising elements to compose 
a general theory of complex systems. 

Nowadays the two major research lines in networks  are the search for 
small world networks in nature and the investigation of  
theoretic models that explain 
the properties of such networks. The first line was pointed out 
in the first paragraph. The second line is dominated by the 
evolving network models \cite{evol1,evol2} and the study of phase order transitions 
in networks \cite{phase1,phase2}. 
This work is situated at an intermediary place between both 
lines. We characterize a network using recent techniques of 
statistical physics but 
instead of looking for real data in the world we construct 
a network using some properties of the set of natural numbers. 

We use as the keystone in the construction of number networks the 
fundamental theorem of number theory which says that each natural 
number has an unique decomposition in prime factors. It means, 
for a number $a$ and prime numbers $p_j$, there is an unique 
product:
\begin{equation}
                  a = p_1^{\alpha_1} p_2^{\alpha_2}... p_m^{\alpha_m}
\label{fund}
\end{equation}
where the exponent $\alpha_j$ is the multiplicity of the prime $p_j$. 
The simplest way to define a link between two vertices (numbers) is the 
following: each time two numbers have a prime in common in equation (\ref{fund})  
they have a connection. We shall see that only this criterium is not 
good enough to construct networks with small clustering coefficient.

This paper has a twofold objective: present a new network model  
based on number properties and characterize this network using 
the approach of modern complex network analyzes. 
In section $2$ we show the model $M$ and its extensions $M_l$, we 
drive our attention to the degree distribution of the network, the set of families and 
the search of invariant quantities. In section $3$ we analyse the distance 
and the clustering coefficient for these  networks. 
 In section $4$ we give our final remarks. 

\section{II. The Model}

In this section we present the  standard model $M$ and its extensions  
$M_l$. We use  the set of natural numbers as vertices 
and an arithmetic property  
 establishes the connections. The connections have no weight  
neither direction. 

\subsection{A. The Basic Model $M$}

In model $M$ the criterium for the existence of a connection between two 
vertices is the following: there is a link between two numbers $a$ and $b$ if 
they share a common divisor. In other words, if $a$ and $b$ have a common 
prime number $p_j$ in decomposition (\ref{fund})  a link is established.  

Figure \ref{fig1} shows a simple realization of this rule for number of 
vertices $N=16$. All the even numbers are interconnected 
(they share the prime $p=2$). Besides the divisors of $3$, $5$, etc, are 
also interconnected.  
The most connected numbers for this $N$ are $6$ and $12$ because they 
are linked to all the even numbers and all the multiples of $3$. 

The first point we explore in the network $M$ is the degree 
distribution, it means the connectivity $k$ of the vertex $i$ 
for vertices ordered according to the connectivity.  
Figure \ref{fig2} displays $k$ versus  $i$. In this 
 figure we have $N=2^{12}=4096$ 
vertices. The vertex $i$ with maximum number of links is $i=2310=2 \times 3 
\times 5 \times 7 \times 11$ which is 
connected with all the	 even numbers 
and with the multiples of $3$, $5$, $7$, and $11$. In general, 
the most connected vertex in a  network $M$ of $N$ vertices corresponds to 
the maximum number $i=p_1 p_2 ... p_m \leq N$ where the $p_j$ 
correspond to the first primes (which are the most connected numbers).  
 In the opposite side of the graphic there 
are the prime numbers $p_j$ that have $k=0$, these numbers  
satisfy the relation $2 p_j \leq N$ (if $p_j$ do not 
fulfill this relation it will be connected to 
the node $2$ and it would not have $k=0$).   
The general view of the degree distribution shown in 
 figure \ref{fig2} is the following. Half of the vertices, which 
corresponds to the even numbers, 
are connected among them, so they have $k \geq N/2$. Otherwise, the odd numbers 
have $k< N/2$ because they are at least not connected to all the even numbers. 
 The frontier between these two sets is indicated by the large 
 plateau that starts at $i=N/2$. 

The plateau in the middle of figure \ref{fig2}  corresponds  
to the family of multiples of the number $2$. In the beginning of this plateau we find the numbers 
$2^{\alpha}$ for $1\leq \alpha \leq 12=log_2 N$. These numbers are connected with all the even 
numbers and only with them. For these numbers $k = N / 2$. The smooth tail 
that comes for $i < N / 2$ is formed by the numbers type $2^{\alpha} p_j \leq N$ 
where $p_j$ is a weak connected prime. These numbers are connected with all the 
even numbers and with the connections of $p_j$ that are just a few. 

The second largest plateau is related to the multiples of three (indicated 
$F_3$ in figure \ref{fig2}).  
One third of the $N$ natural numbers are multiples of three and have connectivity 
larger that all the other odd numbers.  
In the beginning of this plateau we find the 
numbers   of the form $3^{\beta}$ where $1 \leq \beta \leq log_3 N$. 
This plateau is smaller than 
the former one because there are less  multiples of three than two. 
For these numbers $k = N / 3$, it means, they are connected only 
with the multiples of three. The tail of this plateau is formed by 
 the  numbers $3^{\beta} p_j \leq N$, for $p_j$ weakly connected. There are 
also plateaus visible in figure \ref{fig2} 
corresponding to the primes $5$, $7$ and $11$. For these plateaus we have in the 
coordinate axis $ k =N / 5$, 
$N / 7$, and $N / 11$, respectively. 

Another plateau indicated in the figure  is formed by the numbers that 
share the two primes $2$ and $3$, it means the numbers that are multiples of
 $6$. This plateau starts with numbers of the form: $2^{\alpha} 3^{\beta} \leq N$. 
If the network were weighted, $k$ of this plateau would be the sum of $k$ of 
$2$ and $3$ plateaus. As the network does not count  multiple links the 
corresponding value of $k$ is 
smaller than the cited sum. 
 We call 
$F_2$ the even plateau, $F_{10}$ the plateau generated by the numbers 
that are divisible only by $2$ and $5$, and so one. 
The most important family plateaus $F_2$, $F_3$, $F_5$, $F_7$, $F_{10}$, 
$F_{14}$ and $F_{15}$  
are indicated in figure \ref{fig2}. 

Figure \ref{fig2} shows two distinct regions, 
 above $i = N / 2$ the degree distribution  is 
dominated mainly by plateaus of prime numbers of the form: $p_j^{\alpha_j}$ and 
their respective tails. Bellow 
 this number there are only plateaus composed by the combination 
of prime numbers of the form: $p_j^{\alpha_j} p_m^{\alpha_m}$ and their 
tails. In fact in the 
limit of $N \rightarrow \infty$ an infinite number of plateaus would appear in 
the curve whose relative sizes is determined 
by equation (\ref{fund}). We conjecture that in this limit 
a fractal distribution will appear.

We have  $k = N / 2$ for the beginning of the
 plateau $F_2$, and in general     $k = N / p_k$ 
for the families of primes $F_k$. This 
relation suggests an 
interesting scale property: the degrees scales linearly with $N$. We verify 
numerically this fact in figure \ref{fig3} where we plot the normalized connectivity, 
($k/N$),  versus the normalized index,  ($i / N$), 
 for $N=128, 512$ and 
$2048$. This figure verify by simulation that the connectivity of
the model $M$ scales linearly with the number of vertices.

\subsection{B. The extension $M_l$ of the Model $M$}

In the standard number decomposition, equation (\ref{fund}), there is the 
possibility of including, or not, the factor $1$ because all the numbers are 
trivially divisible by the unity. If the number $1$ is 
included in the decomposition all the numbers would share the same common 
divisor and, as a consequence, the network will became trivial: all the 
vertices will be interconnected. In the same way as we excluded the number  
$1$ as a divisor in the criterium for establishing links, we could also exclude   
the number $2$. This idea suggests an alternative procedure 
to define a network model for  natural numbers. 

The network model $M_{p_l}$ is the following. The vertices are again the 
natural numbers and 
the connections are set using equation (\ref{fund}), but we take into account 
only connections of primes $p_j \geq p_l$. In this sense the former model $M$ is 
in fact $M_2$ because the links are established once there is a common factor 
$p_j$ such that  $p_j \geq 2$. In this way, the network $M_3$ take into account 
the primes $p_j=3,5,...$ to establish links, but not the prime $2$.

Figure \ref{fig4} shows the degree distribution ($k$ versus $i$ 
in order of decreasing connectivity) for 
$M_2$, $M_3$, $M_5$, and $M_7$ as indicated in the figure. 
We use in the figure $N=2^{11}$. The curve of $M_2$ is the same curve shown in figure 
\ref{fig2}. The curve of $M_3$ is similar to $M_2$, but the 
largest even plateau, $F_2$, is absent. The largest plateau of $M_3$ 
starts at $i=N/3$, because all the one 
third of the most connected numbers are multiples of $3$. We observe that 
the largest plateau of $M_3$ has the same $k$ of $F_3$ of network $M_2$, in 
fact in both cases the plateaus is formed by the $F_3$ family. 
The curve $M_5$ does 
not show the plateaus corresponding to the families $F_2$ and $F_3$ as expected; 
and in this case the largest plateau is formed by the $F_5$ family. 
The curve of $M_7$ follows the same tendency. 
 As a general trend the curves of degree distribution of 
$M_l$ become smother for increasing $l$ because they have less connections related to 
 important prime numbers and, as a consequence, they present less plateaus. 
The criterium for establishing links in the model $M_l$ 
becomes more restrictive as $l$ increases because the number of 
connections decrease. It is interesting that the average connectivity,
 $< \hspace{-0.1cm} k \hspace{-0.1cm} >= 2 n / N$, normalized by $N$ tends to a constant as 
$N \rightarrow \infty$ ($n$ is the  number of connections in the 
network). In Table $I$ we show $< \hspace{-0.1cm} k \hspace{-0.1cm} > / N$ for the networks 
$M_l$.

{\it \centerline{Table I}}
\begin{tabbing}

\hspace{0.7cm} $M_l$  \hspace{1.8cm} \= $M_2$ \= \hspace{1.2cm} $M_3$ \= \hspace{1.2cm} $M_5$ \= \hspace{1.3cm} $M_7$  \\

\hspace{0.1cm} $< \hspace{-0.1cm} k \hspace{-0.1cm} > / N $  \hspace{1.1cm} \= $0.45$ \=  \hspace{1.1cm} $0.20$ \= \hspace{1.1cm} $0.09$ \= \hspace{1.1cm} $0.05$\\
\hspace{0.7cm} $\bar C$  \hspace{1.8cm} \= $1.78$ \=  \hspace{1.1cm} $3.65$ \= \hspace{1.1cm} $8.22$ \= \hspace{1.1cm} $14.5$\\

\end{tabbing} 

The first conclusion we take from the table is that the model $M$ is not sparse, 
it means, the network does not fulfill the 
condition $< \hspace{-0.1cm} k \hspace{-0.1cm} > \> \ll N$. Therefore  
 we have to be caution to compare properties of such 
graphs with usual complex graphs in the physics literature. 
The reason for $< \hspace{-0.1cm} k \hspace{-0.1cm} > \> \propto N$ is rooted in 
the way connections are established in the network. 
Each time a new even number is added it is connected with, at least, $N/2$ vertices, 
and a multiple of three with $N/3$ vertices. This fact illustrates that in the 
average $< \hspace{-0.1cm} k \hspace{-0.1cm} >$ increases linearly with $N$. 

We also analyze the scaling properties of the model $M_l$. 
Figure \ref{fig5} shows the normalized connectivity, 
($ k / N $),  versus the normalized index,  ($i / N$), for 
the model $M_5$. This curve is similar to the 
one of figure \ref{fig3}, here we also use $N=128,512$ and $2048$. 
The data collapse performed in the figure indicates that 
the network $M_5$ scales with $N$. In fact, this same tendency is 
observed for all networks $M_l$ analyzed. This behavior is related, as before, 
with the plateaus of prime numbers $p_j$ whose connectivity 
scales with $N$. 
The fact that the degree distribution of the networks $M_l$  
scales with $N$ suggests the existence of magnitudes that are independent 
of  $N$. This is the case of $< \hspace{-0.1cm} k \hspace{-0.1cm} > / N$ and 
 this is also the case of the clustering coefficient as we shall see in the 
next section.

\section{III.  Clustering coefficient and network distance} 

In this section we characterize the  network models $M_l$ using the 
distance, $d$, and the clustering coefficient, $C$. One of our objectives in 
this work is to differentiate $M_l$ from random networks, it means, 
networks whose distribution of links among vertices follow 
a Poisson distribution. Therefore we compare 
$d$ and $C$ of $M_l$ with $d$ and $C$ of the random network associated, it means, 
the random network with the same number of vertices and connections.

The distance of a network is defined as the average distance between all the  
two vertices of the network. The clustering coefficient is a global parameter 
$C = \frac{\sum_N c_i}{N}$ which is  based on the local clustering coefficient 
$c_i$. For each vertex $i$ the respective $c_i$, is defined as the 
normalized number of connection among its first neighbors. The parameter $C$ 
measures the average interconnection of the network. Using 
the example of social networks of acquaintances, $c_i$ measures how much 
the friends of a person (vertex $i$) are friends among them. 
For a major treatment on this topic see \cite{barabasi}.

The distance of the random network associated, $d_{rand}$, is 
 estimated by simulation. We note that, because the graph is not sparse, 
 it is not valid that $d \propto ln(N)$. In fact, for 
non-sparse graphs the distance is almost always two 
\cite{mathbook}. Otherwise, the clustering 
coefficient  of the random graph associated, $C_{rand}$,  
is analytically estimated \cite{barabasi} 
and depend only on the number of 
vertices $N$ and the number of connections $n$. For the random graph 
 the clustering coefficient is 
$C_{rand} = < \hspace{-0.1cm} k \hspace{-0.1cm} > / N$. 
We call $\bar d \equiv \frac{d}{d_{rand}}$ 
and $\bar C \equiv \frac{C}{C_{rand}}$ as the normalized distance and clustering 
coefficient.

Figure \ref{fig6} shows the distance $d$ against 
the network size $N$ for the data of the models $M_2$, $M_3$, $M_5$, and $M_7$. 
The graphic is in log-linear form because of the large interval used in $N$. 
 The data confirms the prevision for non-sparse graphs that the 
distance is around $2$.   We estimate $\bar d$ for 
the networks $M_2$, $M_3$, $M_5$, and $M_7$ in the range $2^5 \leq N \leq 2^{12}$ 
and find that $0.5 < \bar d < 0.9$. The general tendency is: 
$\bar d$ slowly increases with $N$. In addition,  $l$ large in models $M_l$ implies  
smaller $\bar d$. This last fact is expected since for large $l$ the connections 
of the network are more selective and organized.

The analysis of $\bar C$ for $M_2$, $M_3$, $M_5$ and $M_7$ shows 
that it increases with $N$ until $N \simeq 2^6$ and stabilize
around a constant value. In Table $II$ it is shown $\bar C$ for 
several $N$ for the model $M_5$, the other models $M_l$ follow a similar trend.

{\it \centerline{Table II}}
\begin{tabbing}
 $\hspace{0.2cm} N$  \hspace{1.2cm} \= 32 \= \hspace{1.5cm} 64 \= \hspace{1.3cm} 128 \= \hspace{1.2cm} 256 \= \hspace{1.1cm} 512 \= \hspace{1.1cm} 1024 \\
 $\hspace{0.2cm} \bar C$  \hspace{1.2cm} \= $6.95$ \=  \hspace{1.1cm} $8.33$ \= \hspace{1.1cm} $8.18$ \= \hspace{1.1cm} $8.19$ \= \hspace{1.1cm}$8.28$ \= \hspace{1.1cm}  $8.36$  \\

 \end{tabbing}

The data points to a constant $\bar C$ in the limit $N 
\rightarrow \infty$. The size invariance of $\bar C$ is compatible 
with the size invariance of the degree distribution. 
The best values of $\bar C$ for $M_2$, $M_3$, $M_5$ and $M_7$ are shown 
in Table $I$. We observe in this table that $\bar C$ for the 
network $M_l$ increases with $l$. The clustering 
coefficient increases as the criterium 
for establishing connections  in the network becomes more selective. 
In fact, for a constant $N$, 
 the number of connection, $n$, in the $M_l$ model decreases   
for high $l$ (see Table $I$). As a result 
$C_{rand} = < \hspace{-0.1cm} k \hspace{-0.1cm} > / N = 2 n / N^2$ decreases 
in contrast with estimated $C$  that remain almost constant.

\section{IV. Final Remarks}
In this work we propose a new network model in which the natural numbers 
are the vertices and the connections are based on their decomposition  
 by prime numbers. Using this criterium we develop a non-sparse 
network ($< \hspace{-0.1cm} k \hspace{-0.1cm}> \sim \ll N$)  which  
have a distance of the order of $2$.    
 If we consider that all the prime numbers 
in the decomposition set a link the network formed is similar to a random graph 
because  the high number of connections implies in a 
small clustering coefficient. If the criterium for establishing 
links becomes more selective, only prime numbers greater than $3$, or $5$, are used 
to establish links the network  has
   a large clustering coefficient. 

We perform data collapse on the data and  verify that the networks 
studied have a degree distribution that is invariant with the number 
of vertices $N$. The general view of the degree distribution is 
a funny discontinuous curve with plateaus of all sizes. 
These plateaus are generated by the families of numbers that share 
the same prime numbers in their decomposition. 

An important class of networks are the scale-free ones. This concept is 
mainly used to distinguish between networks that have exponential and power-law  
degree distributions. In the exponential case 
most of the vertices have a typical connectivity inside a range defined by 
the exponent of the exponential function. 
 In the other side, a power-law case has 
connections of all orders, it does not have a set of vertices with 
 a typical connectivity. The 
 number network $M_l$ do not have a smooth degree distribution 
because of the plateaus formed by the families of prime numbers $F_k$,
therefore it is not possible to fit the degree by 
a smooth curve. On the other side, due to families $F_k$, this network has 
vertices with all orders of connectivity corresponding to all sizes  
of  primes and 
their combinations. In this broad sense the network $M_l$ can be called a 
scale-free network.

This work unfolds an alternative perspective in the study of complex networks. Instead 
of search for real networks in nature we explore deterministic  
 mathematical networks that 
show small distance and high clustering coefficient. Despite the present network 
is non-sparse it is a promising laboratory in the study of degree 
distribution and cluster families.  
In a future work we intend to explore some theoretical developments of this 
problem: an evolving network algorithm for $M_l$ and an analysis of phase 
transition in this model. 

\vspace{0.5cm}

\centerline{\bf Acknowledgments}

The author thanks Marcos Vinicius C. Henriques, Sidney Redner, Liacir S. Lucena, 
and the anonymous referee for the fertile exchange of ideas. 
The author also acknowledges  the financial support of Conselho Nacional
de Desenvolvimento Cient{\'\i}fico e Tecnol{\'o}gico (CNPq)-Brazil and the 
use of the following software for network analysis  
http://vlado.fmf.uni-lj.si/pub/networks/pajek/.


\centerline{FIGURE LEGENDS}

\begin{figure}[ht]
\begin{center}
\caption{ The network of the model $M$ for number of vertices $N=16$. 
The natural numbers are connected according to the prime number decomposition. Two 
numbers have a connection if they share a common prime in the decomposition. }
\label{fig1}
\end{center}
\end{figure}

\begin{figure}[ht]
\begin{center}
\caption{ The connectivity $k$ versus index $i$ for the data of 
network of model $M$. It is 
used $N=4096$. The main families $F_k$ are indicated in the figure. }
\label{fig2}
\end{center}
\end{figure}

\begin{figure}[ht]
\begin{center}
\caption{ The normalized connectivity $k/N$ versus 
normalized index $i/N$ for data of model $M$. 
It is used $N=128$, $512$ and $2048$, as indicated in the figure.  }
\label{fig3}
\end{center}
\end{figure}

\begin{figure}[ht]
\begin{center}
\caption{ The connectivity $k$ versus index $i$ for the models $M_2$, $M_3$, 
$M_5$ and $M_7$ as indicated in the figure. It is used $N=2048$.}
\label{fig4}
\end{center}
\end{figure}

\begin{figure}[ht]
\begin{center}
\caption{ The normalized connectivity $k/N$ versus 
normalized index $i/N$ for the data of model $M_5$. It is used 
 $N=128$, $512$ and $2048$ as indicated in the figure.}
\label{fig5}
\end{center}
\end{figure}

\begin{figure}[ht]
\begin{center}
\caption{The distance $d$ against network size $N$, for 
the models $M_2$, $M_3$, $M_5$, and $M_7$ as indicated in the figure. }
\label{fig6}
\end{center}
\end{figure}

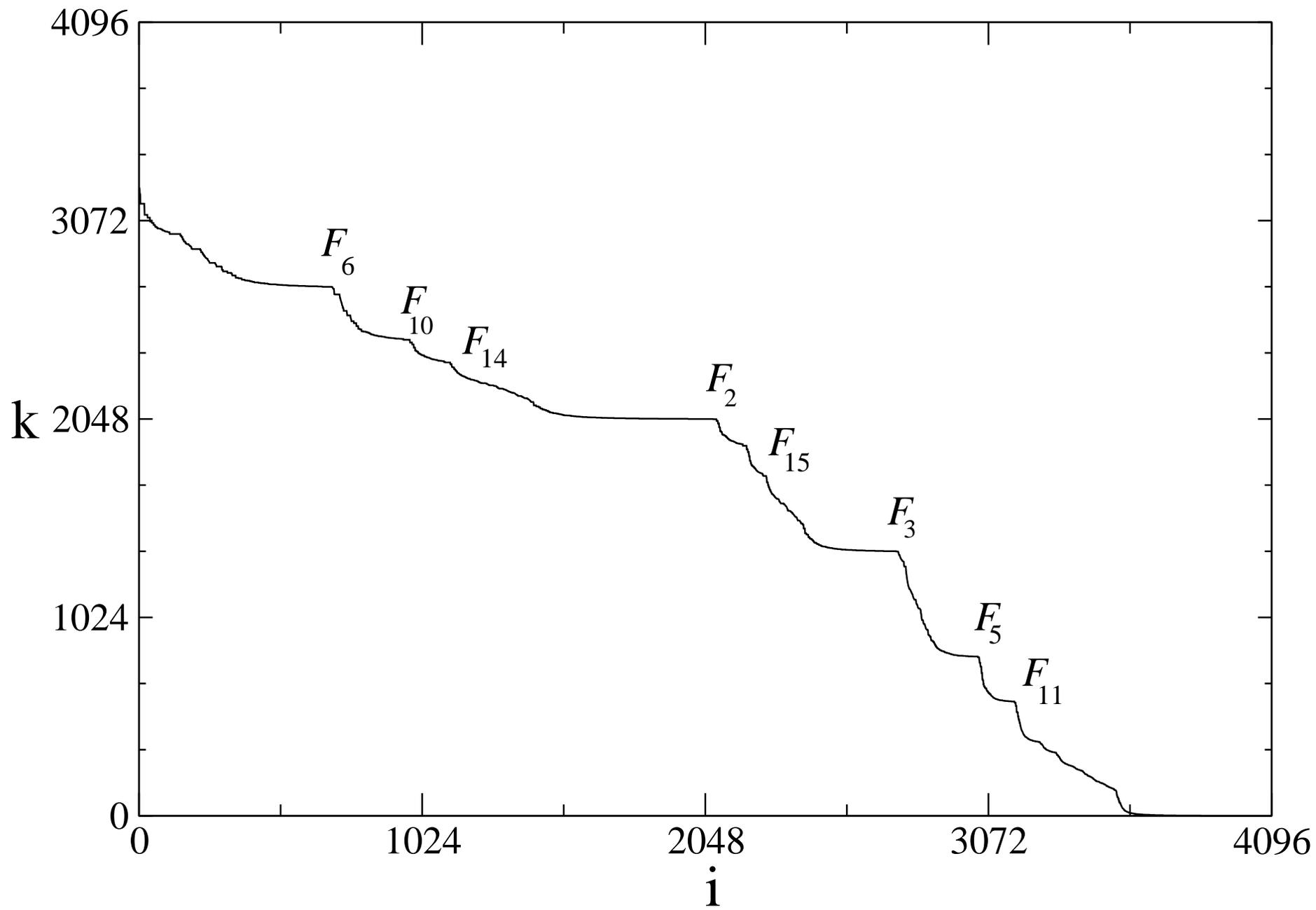
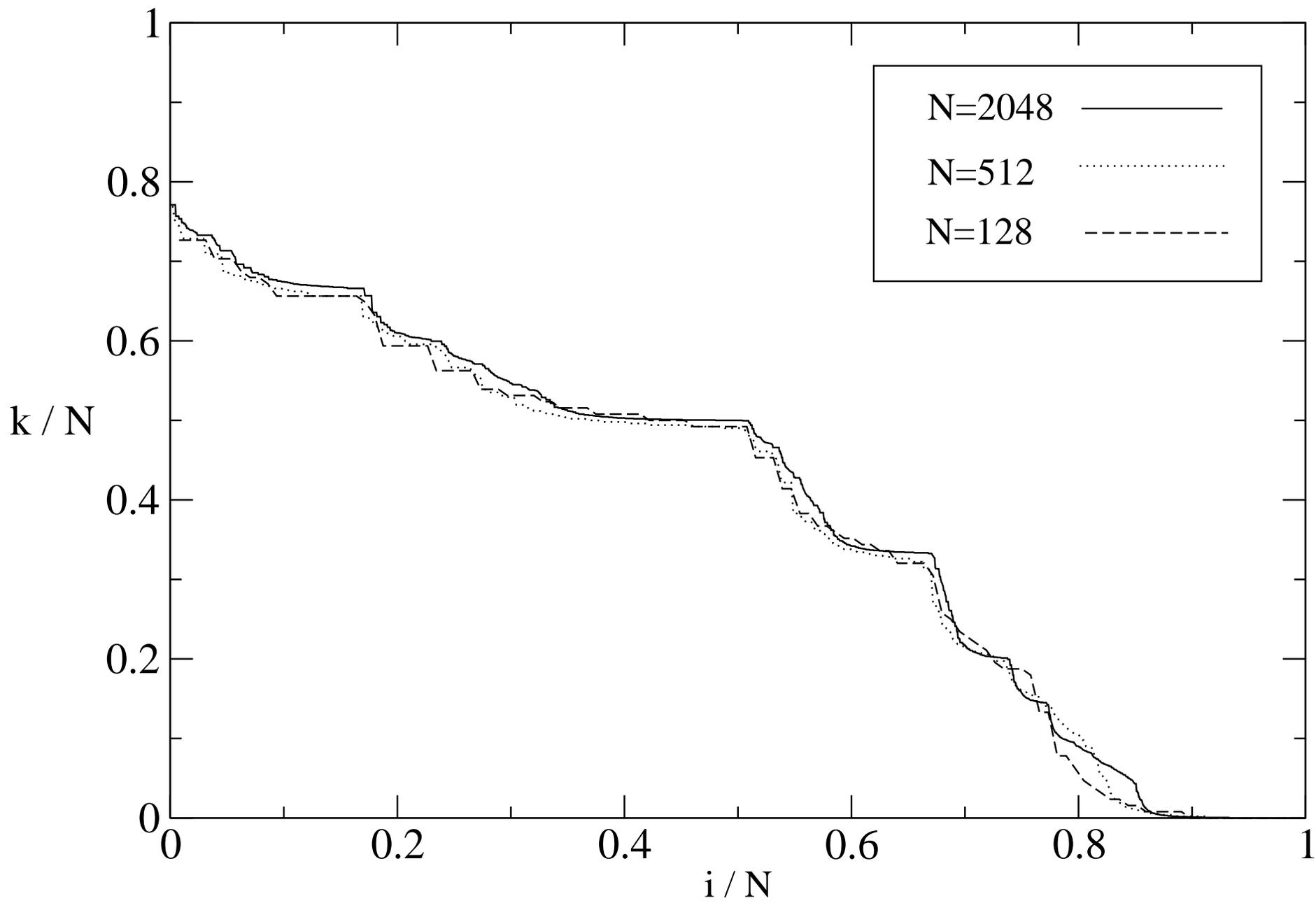
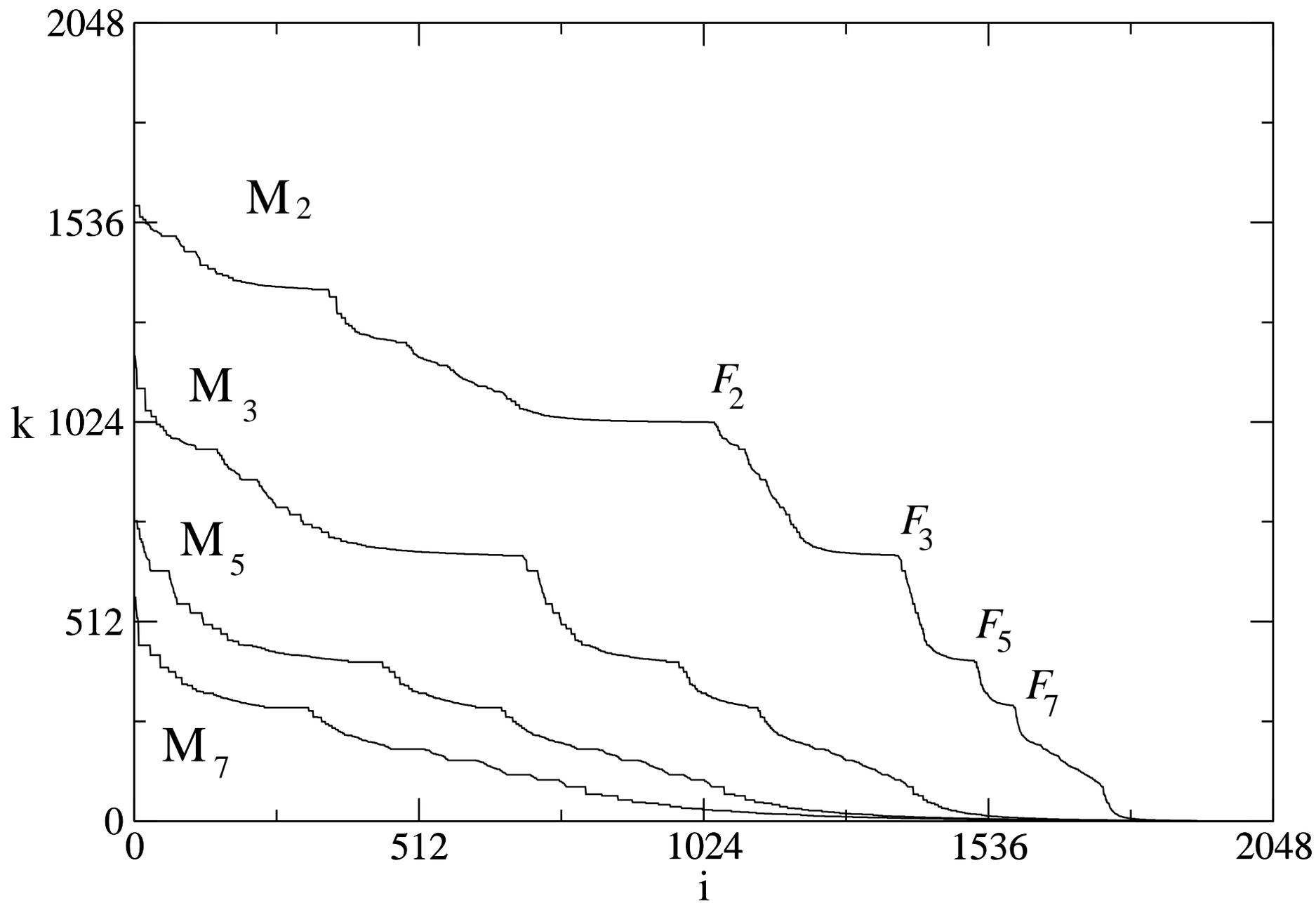
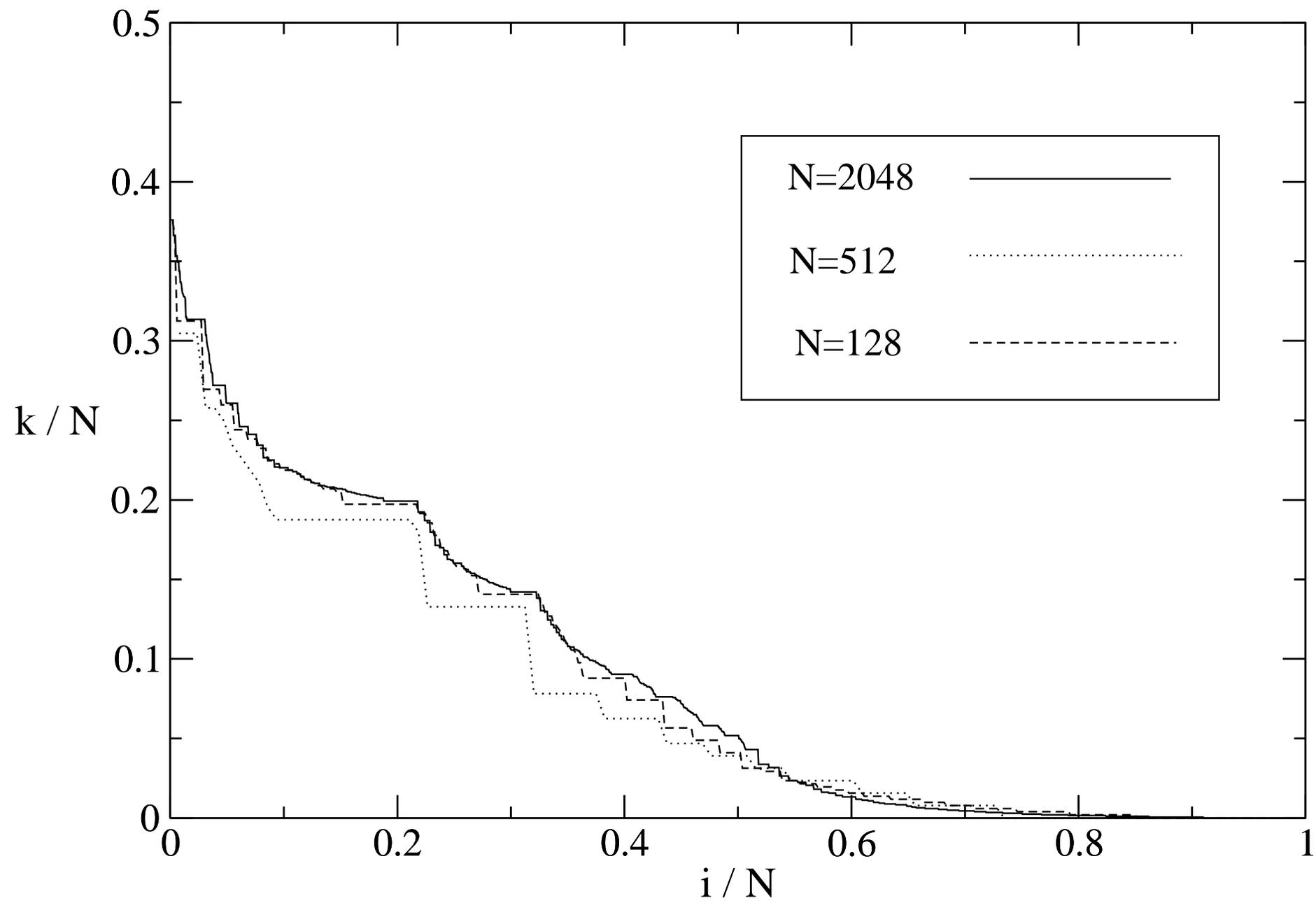
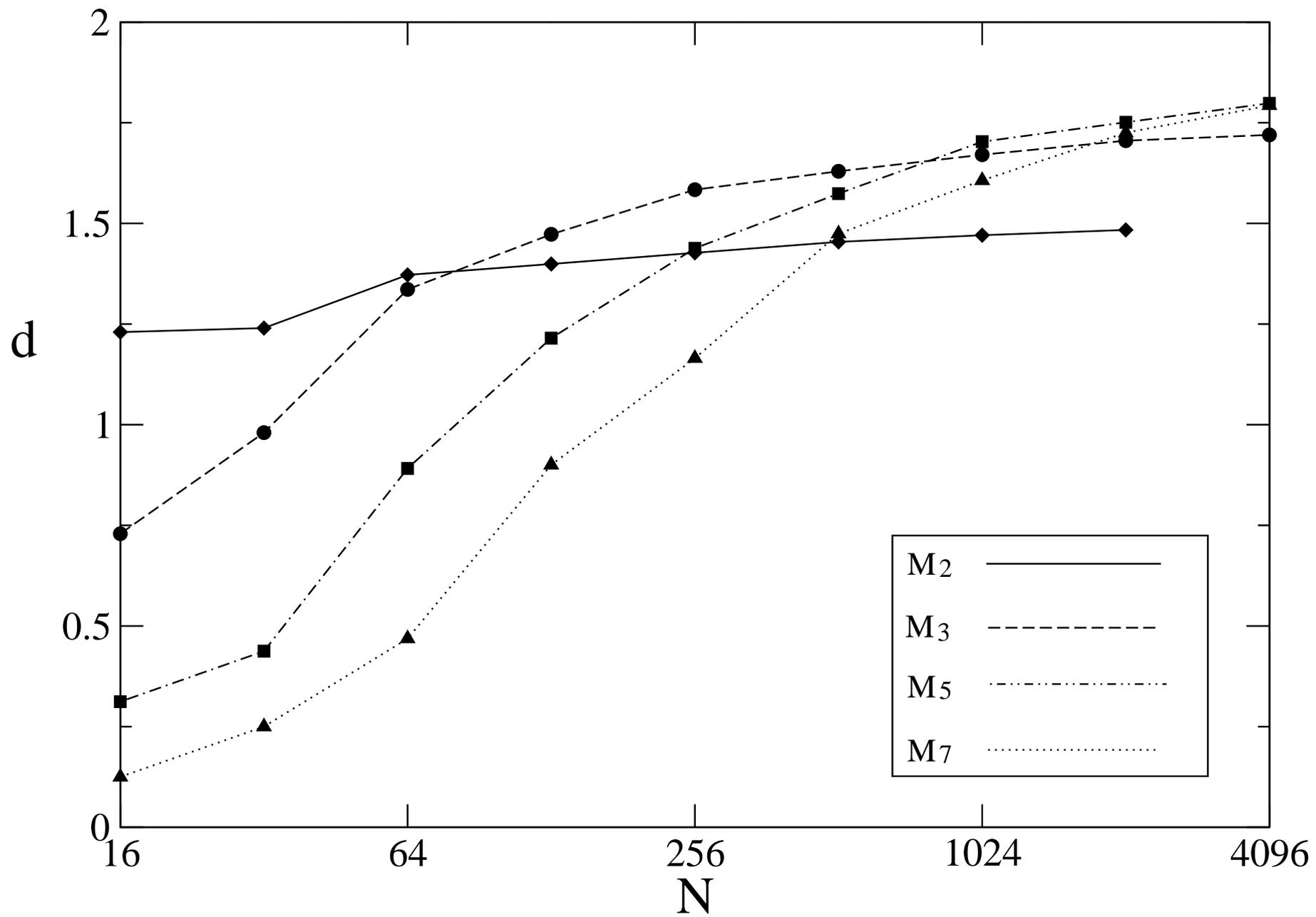
\end{document}